\documentclass[preprint,12pt]{elsarticle}
\usepackage{amsmath}
\usepackage{amssymb,amsfonts,textcomp}
\usepackage{color}
\usepackage{array}
\usepackage{supertabular}
\usepackage{hhline}
\usepackage{hyperref}

\newcommand\textstylest[1]{#1}
\makeatletter
\newcommand\arraybslash{\let\\\@arraycr}
\newcommand\normalsubformula[1]
{\text{\mathversion{normal}$#1$}}

\begin{document}

\begin{frontmatter}

\title{
First-principles investigation of graphitic carbon nitride monolayer with embedded Fe atom}


\author[a,b]{Yusuf Zuntu Abdullahi\corref{cor1}}
\ead{yusufzuntu@gmail.com}

\cortext[cor1]{Corresponding author}
 
\author[a]{Tiem Leong Yoon}
\ead{tlyoon@usm.my} 

\author[a]{Mohd Mahadi Halim}
\author[a]{Md. Roslan Hashim}
\author[c]{Thong Leng Lim}

\address[a]{School of Physics, Universiti Sains Malaysia, 11800 Penang, Malaysia}
\address[b]{Department of Physics, Faculty of Science, Kaduna State University, P.M.B. 2339 Kaduna State, Nigeria}
\address[c]{Faculty of Engineering and Technology, Multimedia University, Jalan Ayer Keroh Lama, 75450 Melaka, Malaysia}







\begin{abstract}
Density-functional theory calculations with spin-polarized generalized gradient approximation and Hubbard $U$ correction is carried out to investigate the mechanical, structural, electronic and magnetic properties of graphitic heptazine with embedded $\mathrm{Fe}$ atom under bi-axial tensile strain and applied perpendicular electric field. It was found that the binding energy of heptazine with embedded $\mathrm{Fe}$ atom system decreases as more tensile strain is applied and increases as more electric field strength is applied. Our calculations also predict a band gap at a peak value of 5 tensile strain but at expense of the structural stability of the system. The band gap opening at 5 tensile strain is due to distortion in the structure caused by the repulsive effect in the cavity between the lone pairs of edge nitrogen atoms and $\mathrm{d}_{{xy}}/\mathrm{d}_{x^2-y^2}$ orbital of Fe atom, hence the unoccupied $\mathrm{p}_z$-orbital is forced to shift towards higher energy. The electronic and magnetic properties of the heptazine with embedded $\mathrm{Fe}$ system under perpendicular electric field up to a peak value of 10 $\mathrm{V/nm}$ is also well preserved despite obvious buckled structure. Such properties may be desirable for diluted magnetic semiconductors, spintronics, and sensing devices.

\end{abstract}
\end{frontmatter}

\section{Introduction} 
Among the nowadays synthesized monolayers, graphitic carbon nitride ($\mathrm{CN}$) sheets have received enormous research interest due to their promising technological advantage such as thermal stability, mechanical, electronic and photocatalytic properties \cite{ref1},\cite{ref2},\cite{ref3},\cite{ref4},\cite{ref5},\cite{ref6},\cite{ref7}.
The $\mathrm{C_3N_4}$ allotrope involves two basic isomers, namely; a graphitic single triazine ($\mathrm{gs}-\mathrm{C_3N_4}$) and graphitic tri-single triazine with formula $\mathrm{gts}-\mathrm{C_3N_4}$ (heptazine) \cite{ref8},\cite{ref9}.
The heptazine has been found to be the most stable structure as compared to other graphitic $\mathrm{CN}$-based allotropes under room temperature \cite{ref10}.
In addition to an intrinsic $\mathrm{sp^3}$ and $\mathrm{sp^2}$ hybridized structure, heptazine possess naturally defined defects which is significant for trapping nanostructures at even dispersions as compared to other monolayer monolayers. Therefore, heptazine could be used as a suitable candidate for photocatalys, sensing and spintronics applications \cite{ref11},\cite{ref12},\cite{ref13},\cite{ref14},\cite{ref15}.
For example, Wang et al. have shown that $\mathrm{g}-\mathrm{C_3N_4}$ doped with $\mathrm{Fe}$ and $\mathrm{Zn}$ exhibits efficient photocalysis more than that of pure $\mathrm{g}-\mathrm{C_3N_4}$ \cite{ref16}.

With respect to spintronics applications, incorporation of magnetic nanostructures into heptazine via atomic doping have recently sparked a renew attention. This is due to the possibility that unique electronic structure such as semiconducting magnetic property and half-metallicity could be tailored in the doped $\mathrm{g}-\mathrm{C_3N_4}$ system. Du {\it et al}. theoretically demonstrated that ferromagnetic ground state with half metallicity can be achieved by uniformly replacing N atom by C atom to form C-doped triazine-based $\mathrm{g}-\mathrm{C_4N_3}$ \cite{ref1}.
However, achieving enduring and stable spin ordering upon incorporation of non-magnetic atoms into the bare monolayer remain uncertain. Recent study has shown an evidence of ferromagnetic behavior under ambient temperature via adsorption of hydrogen dangling bonds at some selected sites on heptazine monolayer \cite{ref17}.
It is hence logical to investigate whether $\mathrm{g-C_3N_4}$ can substantially be magnetized by the well-known method of doping transition metal (TM) atoms into its free-standing form.

By applying first-principles method based on density functional theory (DFT), Zhang \textit{et al.} \cite{ref19}
have confirmed the exotic magnetic behavior of $\mathrm{TMs}@\mathrm{g-CN}$ system obtained in Ref.~\cite{ref18}.
They further reported magnetic exchange coupling of $\mathrm{TMs}@\mathrm{g-CN}$ system which varies dramatically with the separations between the various TM atoms located at the cavity, i.e., from ferromagnetic ($\mathrm{FM}$) to antiferromagnetic ($\mathrm{AFM}$) transition. Moreover, they also show modulated electronic properties for all the heptazine with embedded 3d TM atoms systems. Generally, application of external environments such as electric field, \cite{ref20}
strain, \cite{ref21},\cite{ref23}
and chemical adsorption \cite{ref24},\cite{ref25}
are routinely used for controlling monolayers physical properties. This is because the applications of these external fields would enhance reactivity between the embedded nanostructures and the surrounding atoms arousing stronger interactions. 

Theoretical approach based on $\mathrm{DFT}$ has proved that molybdenum disulfide ($\mathrm{MoS}_2$) monolayer under small deformations would have its band gap decreased. Hence, the semiconducting $\mathrm{MoS}_2$ monolayer exhibits metal electronic character at $8$ deformation \cite{ref26}.
On the iron-doped molybdenum disulfide ($\mathrm{Fe}\text{--}\mathrm{MoS}_2$) system, a magnetic moment is observed to change from  2.04~$\mu _B$ to  4~$\mu _B$ when the biaxial tensile strain reaches $3.5$ \cite{ref21}.
In addition, electronic behaviors, such as magnetic semiconductor and spin-gapless semiconductor are also introduced in $\mathrm{Fe}\text{--}\mathrm{MoS}_2$ system by the strain \cite{ref21}.
In our recent work, it was shown that $\mathrm{Mn}$ atom embedded heptazine band gap increases under small biaxial tensile strain whereas the total magnetic moment remains unchanged \cite{ref22}.
Experimentally, tensile strain is desirable for $\mathrm{CN}$ deposition on a substrate. The straining process of the monolayer during the growth/deposition would result in material's properties modulations \cite{ref27}.
Thus, modulations of the physical properties of $\mathrm{Fe}$ embedded $\mathrm{CN}$ system by an applied external environment present a new research interest (i.e., small deformations and electric field).

In this work, we construct a computational model of heptazine with embedded $\mathrm{Fe}$ atom to theoretically study the ground state properties under small deformations and perpendicular electric field based on $\mathrm{DFT}$ in the presence of  Hubbard $U$. Due to uniform interaction of $\mathrm{Fe}$ atom and the surrounding atoms in the cavity of heptazine, the high spin configuration of the $\mathrm{Fe}$ atom can be preserved. 

\section{Computational method}
All calculations were carried out within the framework of $\mathrm{DFT}+U$ \cite{ref28}
in the QUANTUM ESPRESSO (QE) simulation package \cite{ref29}.
The exchange and correlation energies of strongly localized 3d orbital of TM atoms were treated using Perdew-Burkew-Enzerhof (PBE) of generalized gradient approximation and Hubbard $U$ \cite{ref30}.
The electron-ion core coupling is treated by the Ultrasoft pseudopotentials \cite{ref31}.
For pseudopotentials generation, we refer to the online library of the QE code platform \cite{ref29}.
The cut-off energy for the plain wave basis set used to expand the Kohn-Sham wave functions is found to be 550 $\mathrm{eV}$. Marzari-Vanderbilt smearing with Gaussian spreading was used for integrals to assist in convergences \cite{ref32}.
The Mankhorst-Pack sampling scheme is used for Brillouin zone ($\mathrm{BZ}$) integration. $8\times 8\times 1$ and $15\times 15\times 1$ k-point meshes are chosen for total energy and density of state calculations respectively \cite{ref33}.

We use optimized $2\times 2$ supercell comprising of 56 atoms as shown in the left side of Fig.~\ref{fig1}(c) as our atomic model of pure planar graphitic $\mathrm{gts-C_3N_4}$. Within the atomic structure there are $4$ heptazine units which formed a hexagonal structure with a chemical formula $\mathrm{C_6N_7}$ per unit cell (the right figure in Fig.~\ref{fig1}c show a heptazine unit cell). The $2\times 2$ supercell in the left figure in Fig.~\ref{fig1}(c) is referred as a unit of $\mathrm{GCN}$; while $\mathrm{Fe}$-embedded $\mathrm{GCN}$ is referred as $\mathrm{GCN}-\mathrm{Fe}$. The $\mathrm{GCN}-\mathrm{Fe}$ unit shown in Fig.~\ref{fig1}(a) is the one that was used as the input in our calculation. The purpose of using a $2\times 2$ supercell as in Fig.~\ref{fig1}(a) is to produce negligible interaction between the embedded transition metal atoms. The cavity is indicated by the solid circled line in the $\mathrm{GCN}$ structure Fig.~\ref{fig1}(c). We imposed  a vacuum space of 15 $\text{{\AA}}$ between the periodic images of GCN plane to produce a negligible interaction between the monolayers. All atoms were fully relaxed using \textstylest{Broyden--Fletcher--Goldfarb--Shanno} (BFGS) method until the remaining force on each atom was smaller than 0.05 
$\mathrm{eV}/\text{\AA}$. 

Linear response approach formulated by Cococcioni {\it et al.} \cite{ref35} 
was used to calculate the Hubbard $U$ parameter of the $\mathrm{Fe}$ and 
$\mathrm{Mn}$ atoms embedded in $\mathrm{GCN}$ monolayer. The Hubbard $U$ parameter is extracted from density response function using the following equation: 

\begin{equation}\label{eq1}
\chi _{\normalsubformula{\text{pq}}}=\frac{\partial ^2E}{\partial \alpha _p\alpha _q}=\frac{\partial n_p}{\partial \alpha _q}
\end{equation}
$n_p$ represents the d orbital localized state occupation with respect to site $p$, and $\alpha$ variable is the perturbation potential. Once the density response function is known, the total on-site effective \textit{{U}} value is self-consistently determined and is expressed in equation below,

\begin{equation}\label{eq2}
U_{\mathrm{eff}}=\left(\chi _0^{-1}-\chi ^{-1}\right)
\end{equation}
$\chi$ {and} $\chi _0$ {denote the interacting and non-interacting and density response functions of the system with respect to localized perturbations. The computed Hubbard} $U$ {for} $\mathrm{Fe}$ {atom used in this work is shown in Fig.~\ref{fig2}a.} The obtained  Hubbard $U$  value in this study is more than that adopted by Gosh \textit{et al.} \cite{ref18}
The difference is as a result of distinct bonding environments considered. The $U$ parameter for $\mathrm{Mn}$ in the $\mathrm{GCN}$, which is 3.8 $\mathrm{eV}$, was adopted from our previous calculations \cite{ref22}.

\section{Results and discussions}
\subsection{Structural and mechanical properties of GCN-Fe system}

We first determine the lattice parameters of the optimized $\mathrm{GCN}$ monolayer as illustrated in Fig.~\ref{fig1}a. It can be seen that the optimized lattice parameter is 7.16 $\text{{\AA}}$. The obtained value agrees with the recent report \cite{ref34}.
The bond lengths $l_1$, $l_2$ and $l_{3}$ depicted in Fig.~\ref{fig1}a is consistent with the recent work \cite{ref9}
It is usually recommended to ascertain the robustness of the system. We therefore determine the in-plane elastic moduli of the $\mathrm{GCN}-\mathrm{Fe}$ system from Fig.~\ref{fig2}b. The elastic constants $C_{11},C_{12},$ for uni- and bi-axial strains are 205 N/m and 25.13 N/m respectively. They are calculated from Eq.~\ref{eq3}.
These equations are expressed as:

\begin{eqnarray}\label{eq3}
C_{11}&=&\frac 1{A_0}\left.\left(\frac{{\partial}^2E}{{\partial}s^2}\right)\right|_{s=0} \ \ \ (\mathrm{uni-axial}),
 \nonumber \\
2\left(C_{11}+C_{12}\right)&=&\frac 1{A_0}\left. \left(\frac{{\partial}^2E}{{\partial}s^2}\right)\right|_{s=0} \ \ \ (\mathrm{bi-axial}) 
\end{eqnarray}
where $A_0,E,$ and $s$ are equilibrium unit-cell area, strain energy and applied tensile strain. We then deduced the Young modulus (known as in-plane stiffness), and poison ration from the following equations: \\
In-plane stiffness:
\begin{equation*}
Y=C_{11}\left(1-v^2\right)
\end{equation*}
Poisson's ratio
\begin{equation}\label{eq4}
v=\frac{C_{12}}{C_{11}}
\end{equation}
The calculated in-plane stiffness and Poisson's ratio are $2020.40 \mathrm{GPa}{\cdot}\text{{\AA}} \ (=202.04)$ N/m and $0.12$ respectively. This shows that $\mathrm{GCN}-\mathrm{Fe}$ is slightly weaker in terms of bonding than heptazine \cite{ref3}
and graphyne monolayers \cite{ref36}.
The decreased in stiffness in comparison with pure heptazine can be linked to the increased distortion in the $\mathrm{GCN}$ structure which is caused by the embedded Fe atom in the monolayer. The calculated Poisson's ratio is slightly less than the value for graphene \cite{ref37}.
The bulk modulus is estimated from the well-known method \cite{ref36}
expressed in Eq.~\ref{eq5}.
\begin{equation}\label{eq5}
G=A \times \left.\left(\frac{{\partial}^2E}{{\partial}A^2}\right)\right|_{A_m}
\end{equation}
where $E,A$ and $A_m$ denotes the total strain energy, area of the unit cell, and the equilibrium structure unit cell area respectively. Eq.~\ref{eq5}) (for bulk modulus) stands as the product of equilibrium area and the total energy minimization with respect to the area of $\mathrm{GCN}-\mathrm{Fe}$ system within the linear region in the range of -2 as shown in Fig.~\ref{fig2} (b) iii. Table~\ref{table1} clearly listed the of total strain energy as a function of area derived from the optimized $\mathrm{GCN}-\mathrm{Fe}$ systems lattice constants under small symmetric deformations. An amount of 121.5 N/m is estimated. The estimated value is in order of the value for graphyne monolayer \cite{ref3}
and portrays more hardness compared to the value estimated in our previous work \cite{ref22}.

We also evaluate the stability of $\mathrm{GCN}-\mathrm{Fe}$ system from the binding energy expression as follows

\begin{equation}\label{eq6}
E_b=\left(E_{\mathrm{GCN}}+E_{\mathrm{Fe}}\right)-E_T
\end{equation}
where $E_T,E_{\mathrm{Fe}}$ and $E_{\mathrm{GCN}}$ represent the spin-polarized interacting total energy of the $\mathrm{GCN}-\mathrm{Fe}$ system, the total energy of an isolated $\mathrm{Fe}$ atom and the total energy of pristine GCN respectively. A positive value of $E_b$ indicates more stable $\mathrm{GCN}-\mathrm{Fe}$ system, and a negative value indicates metastable system. Small deformations in terms of tensile strain $s$ can be calculated from the optimized lattice constant $a_0$ of $\mathrm{GCN}-\mathrm{Fe}$ system and it is defined by the formula as $s=\Delta a/a_0$. The average height $h$ difference in $z$-axis of the $\mathrm{Fe}$ atom with respect to all atoms in $\mathrm{GCN}$ monolayer is listed in Table~\ref{table1} as a function of bi-axial strain (positive tensile) for $0-5$ range. 

For the embedded $\mathrm{Fe}$, the computed height $h$ listed in Table~\ref{table2} indicates slight vertical movement of the embedded $\mathrm{Fe}$ atom in the $\mathrm{GCN}$ plane for all deformations. Overall the $\mathrm{GCN}-\mathrm{Fe}$ systems remain nearly planar without much structural distortions after structural relaxation. At a value of  5 tensile strain the structural distortion is more pronounced. This corresponds to a metastable state as evidenced from the computed value of binding energy. We also observe the preservations of the $\mathrm{sp}^2$ structure by measuring the average angle $\theta $ formed by $C$ which is bonded by two neighboring N atoms in the porous site. 

The estimated binding energy as a function of tensile strain obtained from Eq.~\ref{eq6}
shows a decreasing trend. This can be attributed to the small tensile deformations which weaken the bond between the embedded $\mathrm{Fe}$ atom in the porous site and the surrounding atoms. To account for more detailed binding energy modulation, we consider the following variations. For positive elastic moduli, the total energy of the $\mathrm{GCN}-\mathrm{Fe}$ system should increase under small deformation. Correspondingly, the binding energy according to the definition of Eq.~\ref{eq6},
would become lower. We also observe a successive increase in $d_1/d_2$ (see labeled in Fig.~\ref{fig3}c) connecting adjacent heptazine units. The averaged bond length between $\mathrm{Fe}$ atom and N atom in the cavity $d_{\mathrm{Fe}-N}$ also increases as more bi-axial tensile strain is applied. Consequently, we observe gradual separation of $\mathrm{GCN}$ structure into separate units of heptazine. These distortions in the structure resulted in a metastable structure and hence the binding energy decreases. We have also checked the binding energy of $\mathrm{Fe}$ in the $\mathrm{GCN}$ monolayer as a function of applied electric field strength for the range of 0 - 1 V/nm. In contrast, the $\mathrm{GCN}-\mathrm{Fe}$ systems binding energy increases as more electric field strength is applied (see Fig.~\ref{fig3} (c)).

We have also quantitatively determined the $\mathrm{GCN}-\mathrm{Fe}$ system magnetic moment per unit cell and charge transfer from the $\mathrm{Fe}$ atom into the $\mathrm{GCN}$ monolayer based on  {Lowdin's charge analysis} \cite{ref38}
The unstrained $\mathrm{GCN}-\mathrm{Fe}$ system (i.e. @  0 tensile deformation) magnetic moment tallies well with the previous literature \cite{ref18}.
However, the results show that the magnetic moment is less sensitive to the small symmetric tensile deformation, i.e. the magnetic moments for different strained system have not deviate from unstrained system. This results show that if $\mathrm{Fe}$ atoms located in all the cavities of $\mathrm{GCN}$ monolayer, its isolated value of the magnetic moment may be well retained. Such property is needed for future spintronic practical applications.  

We found that the high spin configuration of $\mathrm{Fe}$ atom do not deviate from the isolated ones. Hence the magnetic moment of the $\mathrm{Fe}$ atom is relatively preserved. Moreover, we observe lesser charge redistribution of within the sub-orbitals of the embedded $\mathrm{Fe}$ atom as captured in the {Lowdin's charge analysis}. The calculated charge transfer $Q$ from the embedded $\mathrm{Fe}$ atom to the $\mathrm{GCN}$ monolayer as shown in Table~\ref{table1} confirms an ionic bonding behavior between the $\mathrm{Fe}$ atom and the $\mathrm{GCN}$ monolayer for all systems. The charge localization around N atoms in the charge-density difference plot of unstrained system in Fig.~\ref{fig1}(d) further testifies the ionic interaction. The dominant features of charge accumulation around the six surrounding N atoms (orange color 4-point starts) atoms is due to its superior electronegative. There is also an evidence of charge exchange between the atoms in the porous site. These excess charge depletions portray covalent bonding character.

To facilitate a better explanation of the electronic property modulation of $\mathrm{GCN}$ when single Fe atom is embedded on its surface, we first look at the unstrained total density of state (TDOS) of the pristine GCN [see Fig.~\ref{fig3}(a)]. Our result confirms the previously reported nonmagnetic and semiconducting behavior of pristine $\mathrm{GCN}$ system \cite{ref9},\cite{ref39}.
 It is well known that the covalent hybridization of any 2D monolayer orbitals and the orbitals of the foreign atom play a major role in defining the electronic property \cite{ref40},\cite{ref41}.
Upon doping with $\mathrm{Fe}$ atoms, the electronic property of pristine $\mathrm{GCN}$ is clearly modulated. This is obviously seen by comparing Fig.~\ref{fig4}(e) [which shows  the spin-polarized TDOS for the equilibrium structure of GCN-Fe system for the unstrained ($s=0$) case] and Fig.~\ref{fig3}(a) (which shows the TDOS of the pristine monolayer). This electronic behavior confirms the structural distortion of $\mathrm{GCN}$ upon $\mathrm{TM}$ atom embedment. It can equally be observed a maintained intrinsic semiconducting property of $\mathrm{GCN}$ in the vicinity of Fermi level. The asymmetric TDOS clearly indicates the presence of magnetic moment in the system.

Referring to Fig.~\ref{fig4}, the atomic orbital contributions to valence and conduction bands that lead to the electronic property modulation can be analyzed. To understand the pattern of the orbital hybridization between the $\mathrm{Fe}$ and six surrounding Natoms (N$_{6\mathrm{EG}}$), we have plotted the $\mathrm{PDOS}$ of $\mathrm{sp}$ and d orbitals of an isolated $\mathrm{Fe}$ atom (Hubbard $U$ not included) in Figs.~\ref{fig4}(a), (b). By comparing Figs.~\ref{fig4}(a), (b) and Figs.~\ref{fig4}(f) ii, iii (for that of $\mathrm{Fe}$ in $\mathrm{GCN}$), it is found that the pattern of orbital distributions are relatively maintained. This shows that the hybridization between the orbitals of $\mathrm{Fe}$ atom in $\mathrm{CGN}$ with the surrounding atoms does not lead to major orbital redistribution. Figs.~\ref{fig4}(f)i-iii reveals the $\mathrm{PDOS}$ contributions of the embedded $\mathrm{Fe}$ atom and the N$_{6\mathrm{EG}}$ orbitals which are sp-hybridized. The N$_{6\mathrm{EG}}$ are the nearest neighboring N atoms in the porous site, and formed covalent bond with the $\mathrm{Fe}$ atom. It is well known that the covalent hybridization of any 2D monolayer orbitals and the orbitals of the foreign atom play a major role in defining the electronic property \cite{ref40},\cite{ref41}.
At the bottom of the conduction band, there is obvious dominant features of $\mathrm{p}_z$-like orbitals of the N\textsubscript{6EG} in both spin states; whereas the top of the valence band is formed by the $\sigma $-like orbitals including lone pairs [see $\mathrm{PDOS}$ in Figs.~\ref{fig4}(f) i-iii].

The $\sigma $-like orbitals contributing to the bonding are largely occupied by sp-like orbitals of the N$_{6\mathrm{EG}}$ and  s-, $\mathrm{p}_x$-like like orbitals of the Fe atom around  -6.5 $\mathrm{eV}$ and -8.0 $\mathrm{eV}$. Around  -4.5 $\mathrm{eV}$ there is a covalent bonding which is occupied by sp-like orbitals of the N$_{6\mathrm{EG}}$ as well as  s-, $\mathrm{d}_{zx}$- and $\mathrm{d}_{z^2}$- like orbitals of the $\mathrm{Fe}$ atom in the minority spin state. The proportion of the dominant orbitals in the majority spin states are mainly from $\sigma $-like orbitals orbitals of the N\textsubscript{6EG}, with small $\mathrm{sp}$-like orbitals promotion by the $\mathrm{Fe}$ atom. The rest of the sp-like orbitals including lone pairs of the N$_{6\mathrm{EG}}$ contribute to the planar geometry. This is obvious at approximately -0.12 $\mathrm{eV}$, where both spin up and spin down channels is contributed by $\mathrm{p}_y$-, $\mathrm{p}_x$-like orbitals of the N$_{6\mathrm{EG}}$. It also follows that the $\mathrm{sp}$-likes orbitals of the $\mathrm{Fe}$ atom in both spin up and spin down channels confirms intra-orbitals electron transfer. 

On the intrinsic band gap, we observe a linear state degeneracy to the left of Fermi level and at approximately  0.04 $\mathrm{eV}$ in both spin channels. The TDOS plots and band structures in Fig.~\ref{fig4}(c), (d) clarify this degeneracy. The minimum conduction band of both spin up and spin down channels lies on the Fermi level lines. The $\mathrm{GCN}-\mathrm{Fe}$ also system maintains its metallic properties under biaxial tensile strain in the range  0 - 4. The semiconducting property becomes pronounced at a maximum value of  5 tensile strain. This is due to distortion which causes delocalization $\mathrm{p}_z$- orbital at the bottom of the conduction band. The distortion in the structure is as a result of the repulsive effect in the cavity between the lone pairs of edge nitrogen atoms, the $\mathrm{p}_x$, $\mathrm{p}_y$ orbitals of N and $\mathrm{d}_{{xy}}$, $\mathrm{d}_{x^2-y^2}$ of $\mathrm{Fe}$. Consequently, the lone pairs of  N-($\mathrm{p}_x$, $\mathrm{p}_y$) orbitals become misaligned in trying to reduce the steric repulsion which makes $\mathrm{p}_z$- to shift towards higher energy, hence opening the band gap at a peak tensile strain as depicted in Fig.~\ref{fig3}(b). This happens at the expense of the stability of the system [see Table~\ref{table1}]. 

\subsection{Electric field effect}

To explore other effects of an applied external environment on the electronic property, the perpendicular electronic field is applied to the $\mathrm{GCN}-\mathrm{Fe}$ system. As mentioned in the first part, application of external fields has been found to be an effective way to controlled electronic property of 2D materials \cite{ref20},\cite{ref21},\cite{ref22},\cite{ref24}.
We have carried out calculations with an electric field perpendicular to the $\mathrm{GCN}-\mathrm{Fe}$ monolayer ranging from $0.0$. The geometry was fully-optimized under each applied electric field strength. As a result of an applied electric field strength, the structure buckles. The buckling of $\mathrm{GCN}-\mathrm{Fe}$ and $\mathrm{GCN}-\mathrm{Mn}$ (for comparison) systems is due to distortion caused by $\mathrm{d}_{{xy}}$, $\mathrm{d}_{x^2-y^2}$ of $\mathrm{Fe}$ in trying to lower the system's energy (see right Fig.~\ref{fig1}(d)). Orbital hybridizations between $\mathrm{p}_x$, $\mathrm{p}_y$ orbitals of N and $\mathrm{d}_{{xy}}$, $\mathrm{d}_{x^2-y^2}$ of $\mathrm{Fe}$, which lie on the same plane are clearly seen in the projected density of states. This orbital mixed supports the picture depicted above. Despite the obvious wrinkles, the $\mathrm{GCN}-\mathrm{Mn}$ and $\mathrm{GCN}-\mathrm{Fe}$ systems remain semiconducting and metallic respectively [see Fig.~\ref{fig5}(a)-(c)]. 

Furthermore, this confirms that $\mathrm{GCN}$ monolayer displays no exception to the formation of buckled structure due to interplay in the N atom with $\mathrm{sp}^3$ and $\mathrm{sp}^2$ hybridized structure. Application of electric field promotes more repulsive effects between lone pairs, $\mathrm{p}_x$, $\mathrm{p}_y$ orbitals of N and $\mathrm{d}_{{xy}}$, $\mathrm{d}_{x^2-y^2}$ of $\mathrm{Fe}$ as a result of orbitals hybridizations. Moreover, the magnetic moment per unit cell obtained in these calculations doesn't change much with the applied electric field [see Fig.~\ref{fig3}(d)]. These conditions of having preserved electronic and magnetic moment properties under applied electronic field of both $\mathrm{GCN}-\mathrm{Fe}$ and $\mathrm{GCN}-\mathrm{Mn}$ could be used in sensing/actuating applications

\section{Conclusions}
In summary, using spin-polarized $\mathrm{DFT}+U$ calculations, the structural, mechanical, electronic and magnetic properties of $\mathrm{GCN}-\mathrm{Fe}$ system under small tensile deformation and perpendicular electronic field are investigated. It is found that the $\mathrm{GCN}-\mathrm{Fe}$ structure is mechanically stable. Based on the structural feature distortions of the monolayer, the binding energy of $\mathrm{GCN}-\mathrm{Fe}$ system is also found to uniformly decrease. Consequently, the average  N-C-N bond angles within the cavity increases in trying to retain its $\mathrm{sp}^2$ hybridize structure. The electronic character of different strained $\mathrm{GCN}-\mathrm{Fe}$ systems shows semiconducting and metallic behaviors. The magnetic moment of the $\mathrm{GCN}-\mathrm{Fe}$ system is maintained for all deformations. The electronic property modulation can be related to the orbital hybridization in the $\mathrm{GCN}-\mathrm{Fe}$ system. The distortion in the structure as a result of repulsive effect in the cavity between the lone pairs of edge nitrogen atoms and $\mathrm{d}_{{xy}}/\mathrm{d}_{x^2-y^2}$ orbitals lead to the band gap opening at $5$ tensile strain. It is also shown that there is no change in the electronic and magnetic properties of the $\mathrm{GCN}$-Fe system under electric field up to a peak value of  10 $\mathrm{V}/\mathrm{nm}$ in spite of the obvious buckled structure. Our calculations present theoretical insights of the considered system for future applications in spintronics and sensing/actuating devices.

\section*{Acknowledgments}
T. L. Yoon wishes to acknowledge the support of Universiti Sains Malaysia RU grant (No. 1001/PFIZIK/811240). Figures
showing atomic model and 2D charge-density difference plots are generated using the XCRYSDEN program Ref.~\cite{Kokalj:CMS2003}. 
We gladfully acknowledge Dr. Chan Huah Yong from USM School of Computer Science, and Prof. Mohd. Zubir Mat Jafri from USM School of Physics, for providing us computing resources to carry out part of the calculations done in this paper.

\bibliographystyle{elsarticle-num}
\bibliography{ref2ndpaper}

\begin{thebibliography}{10}
\expandafter\ifx\csname url\endcsname\relax
  \def\url#1{\texttt{#1}}\fi
\expandafter\ifx\csname urlprefix\endcsname\relax\def\urlprefix{URL }\fi
\expandafter\ifx\csname href\endcsname\relax
  \def\href#1#2{#2} \def\path#1{#1}\fi

\bibitem{ref1}
A.~Du, S.~Sanvito, S.~C. Smith, First-principles prediction of metal-free
  magnetism and intrinsic half-metallicity in graphitic carbon nitride,
  Physical review letters 108~(19) (2012) 197207.

\bibitem{ref2}
M.~H.~V. Huynh, M.~A. Hiskey, J.~G. Archuleta, E.~L. Roemer, Preparation of
  nitrogen‐rich nanolayered, nanoclustered, and nanodendritic carbon
  nitrides, Angewandte Chemie 117~(5) (2005) 747--749.

\bibitem{ref3}
Y.~Z. Abdullahi, T.~L. Yoon, M.~M. Halim, M.~R. Hashim, T.~L. Lim, Mechanical
  and electronic properties of graphitic carbon nitride sheet: First-principles
  calculations, Solid State Communications 248 (2016) 144--150.

\bibitem{ref4}
P.~Niu, G.~Liu, H.-M. Cheng, Nitrogen vacancy-promoted photocatalytic activity
  of graphitic carbon nitride, The Journal of Physical Chemistry C 116~(20)
  (2012) 11013--11018.

\bibitem{ref5}
P.~Niu, L.~Zhang, G.~Liu, H.-M. Cheng, Graphene‐like carbon nitride
  nanosheets for improved photocatalytic activities, Advanced Functional
  Materials 22~(22) (2012) 4763--4770.

\bibitem{ref6}
X.~Wang, S.~Blechert, M.~Antonietti, Polymeric graphitic carbon nitride for
  heterogeneous photocatalysis, Acs Catalysis 2~(8) (2012) 1596--1606.

\bibitem{ref7}
X.~Li, J.~Zhou, Q.~Wang, Y.~Kawazoe, P.~Jena, Patterning graphitic {C--N}
  sheets into a kagome lattice for magnetic materials, The journal of physical
  chemistry letters 4~(2) (2012) 259--263.

\bibitem{ref8}
V.~N. Khabashesku, J.~L. Zimmerman, J.~L. Margrave, Powder synthesis and
  characterization of amorphous carbon nitride, Chemistry of materials 12~(11)
  (2000) 3264--3270.

\bibitem{ref9}
G.~Algara-Siller, N.~Severin, S.~Y. Chong, T.~Björkman, R.~G. Palgrave,
  A.~Laybourn, M.~Antonietti, Y.~Z. Khimyak, A.~V. Krasheninnikov, J.~P. Rabe,
  Triazine-based graphitic carbon nitride: A two-dimensional semiconductor,
  Angewandte Chemie 126~(29) (2014) 7580--7585.

\bibitem{ref10}
D.~M. Teter, R.~J. Hemley, Low-compressibility carbon nitrides, Angewandte
  Chemie 271~(5245) (1996) 53--55.

\bibitem{ref11}
Q.~H. Wang, K.~Kalantar-Zadeh, A.~Kis, J.~N. Coleman, M.~S. Strano, Electronics
  and optoelectronics of two-dimensional transition metal dichalcogenides,
  Nature nanotechnology 7~(11) (2012) 699--712.

\bibitem{ref12}
X.-H. Li, X.~Wang, M.~Antonietti, Mesoporous g-{C3N4} nanorods as
  multifunctional supports of ultrafine metal nanoparticles: Hydrogen
  generation from water and reduction of nitrophenol with tandem catalysis in
  one step, Chemical Science 3~(6) (2012) 2170--2174.

\bibitem{ref13}
J.~Zhang, M.~Grzelczak, Y.~Hou, K.~Maeda, K.~Domen, X.~Fu, M.~Antonietti,
  X.~Wang, Photocatalytic oxidation of water by polymeric carbon nitride
  nanohybrids made of sustainable elements, Chemical Science 3~(2) (2012)
  443--446.

\bibitem{ref14}
X.~Zhang, X.~Xie, H.~Wang, J.~Zhang, B.~Pan, Y.~Xie, Enhanced photoresponsive
  ultrathin graphitic-phase {C3N4} nanosheets for bioimaging, Journal of the
  American Chemical Society 135~(1) (2012) 18--21.

\bibitem{ref15}
X.~Zhang, H.~Wang, H.~Wang, Q.~Zhang, J.~Xie, Y.~Tian, J.~Wang, Y.~Xie,
  Single-layered graphitic-{C}3{N}4 quantum dots for two‐photon fluorescence
  imaging of cellular nucleus, Advanced Materials 26~(26) (2014) 4438--4443.

\bibitem{ref16}
X.~Wang, X.~Chen, A.~Thomas, X.~Fu, M.~Antonietti, Metal-containing carbon
  nitride compounds: A new functional organic--metal hybrid material, Advanced
  Materials 21~(16) (2009) 1609--1612.

\bibitem{ref17}
K.~Xu, X.~Li, P.~Chen, D.~Zhou, C.~Wu, Y.~Guo, L.~Zhang, J.~Zhao, X.~Wu,
  Y.~Xie, Hydrogen dangling bonds induce ferromagnetism in two-dimensional
  metal-free graphitic-{C3N4} nanosheets, Chemical Science 6~(1) (2015)
  283--287.

\bibitem{ref19}
S.~Zhang, R.~Chi, C.~Li, Y.~Jia, Structural, electronic and magnetic properties
  of 3d transition metals embedded graphene-like carbon nitride sheet: {A}
  {DFT}+{{\it U}} study, Physics Letters A 380~(14) (2016) 1373--1377.

\bibitem{ref18}
D.~Ghosh, G.~Periyasamy, B.~Pandey, S.~K. Pati, Computational studies on
  magnetism and the optical properties of transition metal embedded graphitic
  carbon nitride sheets, Journal of Materials Chemistry C 2~(37) (2014)
  7943--7951.

\bibitem{ref20}
B.~Huang, H.~Xiang, J.~Yu, S.-H. Wei, Effective control of the charge and
  magnetic states of transition-metal atoms on single-layer boron nitride,
  Physical review letters 108~(20) (2012) 206802.

\bibitem{ref21}
Z.~Chen, J.~He, P.~Zhou, J.~Na, L.~Sun, Strain control of the electronic
  structures, magnetic states, and magnetic anisotropy of {F}e doped
  single-layer {MoS$_2$}, Computational Materials Science 110 (2015) 102--108.

\bibitem{ref23}
J.~Qi, X.~Li, X.~Chen, K.~Hu, Strain tuning of magnetism in {M}n doped
  {MoS$_2$} monolayer, Journal of Physics: Condensed Matter 26~(25) (2014)
  256003.

\bibitem{ref24}
Y.~Z. Abdullahi, M.~M. Rahman, A.~Shuaibu, S.~Abubakar, H.~Zainuddin,
  R.~Muhida, H.~Setiyanto, Density functional study of manganese atom
  adsorption on hydrogen-terminated armchair boron nitride nanoribbons, Physica
  B: Condensed Matter 447 (2014) 65--69.

\bibitem{ref25}
F.~Ersan, O.~Arslanalp, G.~Gokoglu, E.~Akturk, Effect of adatoms and molecules
  on the physical properties of platinum-doped and-substituted silicene: A
  first-principles investigation, Applied Surface Science 371 (2016) 314--321.

\bibitem{ref26}
E.~Scalise, M.~Houssa, G.~Pourtois, V.~Afanas'ev, A.~Stesmans, Strain-induced
  semiconductor to metal transition in the two-dimensional honeycomb structure
  of {MoS$_2$}, Nano Research 5~(1) (2012) 43--48.

\bibitem{ref22}
Y.~Z. Abdullahi, T.~L. Yoon, M.~M. Halim, M.~R. Hashim, M.~Z.~M. Jafri, L.~T.
  Leng, Geometric and electric properties of graphitic carbon nitride sheet
  with embedded single manganese atom under bi-axial tensile strain, Current
  Applied Physics 16~(8) (2016) 809--815.

\bibitem{ref27}
S.~Zuluaga, L.-H. Liu, N.~Shafiq, S.~M. Rupich, J.-F. Veyan, Y.~J. Chabal,
  T.~Thonhauser, Structural band-gap tuning in g-{C3N4}, Physical Chemistry
  Chemical Physics 17~(2) (2015) 957--962.

\bibitem{ref28}
P.~Hohenberg, W.~Kohn, Inhomogeneous electron gas, Physical review 136~(3B)
  (1964) B864.

\bibitem{ref29}
P.~Giannozzi, S.~Baroni, N.~Bonini, M.~Calandra, R.~Car, C.~Cavazzoni,
  D.~Ceresoli, G.~L. Chiarotti, M.~Cococcioni, I.~Dabo, {QUANTUM ESPRESSO}: A
  modular and open-source software project for quantum simulations of
  materials, Journal of Physics: Condensed Matter 21~(39) (2009) 395502.

\bibitem{ref30}
J.~P. Perdew, K.~Burke, M.~Ernzerhof, Generalized gradient approximation made
  simple, Physical review letters 77~(18) (1996) 3865.

\bibitem{ref31}
D.~Vanderbilt, Soft self-consistent pseudopotentials in a generalized
  eigenvalue formalism, Physical Review B 41~(11) (1990) 7892.

\bibitem{ref32}
N.~Marzari, D.~Vanderbilt, A.~De~Vita, M.~Payne, Thermal contraction and
  disordering of the {A}l (110) surface, Physical review letters 82~(16) (1999)
  3296.

\bibitem{ref33}
H.~J. Monkhorst, J.~D. Pack, Special points for brillouin-zone integrations,
  Physical Review B 13~(12) (1976) 5188.

\bibitem{ref35}
M.~Cococcioni, S.~De~Gironcoli, Linear response approach to the calculation of
  the effective interaction parameters in the {LDA}+{{\it U}} method, Physical
  Review B 71~(3) (2005) 035105.

\bibitem{ref34}
S.~M. Aspera, H.~Kasai, H.~Kawai, Density functional theory-based analysis on
  {O$_2$} molecular interaction with the tri-s-triazine-based graphitic carbon
  nitride, Surface Science 606~(11) (2012) 892--901.

\bibitem{ref36}
M.~Asadpour, S.~Malakpour, M.~Faghihnasiri, B.~Taghipour, Mechanical properties
  of two-dimensional graphyne sheet, analogous system of {BN} sheet and
  graphyne-like {BN} sheet, Solid State Communications 212 (2015) 46--52.

\bibitem{ref37}
E.~Cadelano, P.~L. Palla, S.~Giordano, L.~Colombo, Elastic properties of
  hydrogenated graphene, Physical Review B 82~(23) (2010) 235414.

\bibitem{ref38}
P.-O. Löwdin, On the non‐orthogonality problem connected with the use of
  atomic wave functions in the theory of molecules and crystals, The Journal of
  Chemical Physics 18~(3) (1950) 365--375.

\bibitem{ref39}
T.~Wei, Wei~Jacob, Strong excitonic effects in the optical properties of
  graphitic carbon nitride g-{C3N4} from first principles, Physical Review B
  87~(8) (2013) 085202.

\bibitem{ref40}
Z.~Chen, J.~He, P.~Zhou, J.~Na, L.~Sun, Strain control of the electronic
  structures, magnetic states, and magnetic anisotropy of {F}e doped
  single-layer {MoS$_2$}, Computational Materials Science 110 (2015) 102--108.

\bibitem{ref41}
M.~M. Rahman, Y.~Z. Abdullahi, A.~Shuaibu, S.~Abubakar, H.~Zainuddin,
  R.~Muhida, H.~Setiyanto, Density functional study of structural stabilities,
  electric and magnetic properties of vanadium adsorption on graphene, Journal
  of Computational and Theoretical Nanoscience 12 (2015) 1995--2002.

\bibitem{Kokalj:CMS2003}
A.~Kokalj, Computer graphics and graphical user interfaces as tools in
  simulations of matter at the atomic scale, Computational Materials Science
  28~(2) (2003) 155--168.

\end{thebibliography}

\begin{table}[p]
\caption{Computed optimized lattice parameters, unit cell area and total energy of $\mathrm{GCN}-\mathrm{Fe}$ monolayer for elastic moduli measurement.}
\label{table1}
\begin{center}
\begin{supertabular}{m{0.5in}m{0.5in}m{1.4031599in}m{1.1302599in}m{1.3462598in}}
\hline \\
{\centering Strain\par}

\centering (\%) &
{\centering Area\par}

\centering ({\AA}\textsuperscript{2}) &
{\centering Total energy\par}

{\centering Bi-axial\par}

\centering (Ry) &
{\centering Total energy\par}

{\centering Uni-axial\par}

\centering (Ry) &
{\centering Lattice parameter\par}

\centering\arraybslash ({\AA})\\\hline
\centering -0.02 &
\centering 170.73 &
\centering -1157.87112 &
\centering -1157.81623 &
\centering\arraybslash 14.04\\
\centering -0.015 &
\centering 172.42 &
\centering -1157.90539 &
\centering -1157.83177 &
\centering\arraybslash 14.11\\
\centering -0.01 &
\centering 174.13 &
\centering -1157.92871 &
\centering -1157.84249 &
\centering\arraybslash 14.18\\
\centering -0.005 &
\centering 175.84 &
\centering -1157.94132 &
\centering -1157.84849 &
\centering\arraybslash 14.25\\
\centering 0 &
\centering 177.64 &
\centering -1157.94375 &
\centering -1157.84931 &
\centering\arraybslash 14.32\\
\centering 0.005 &
\centering 179.37 &
\ \ \ \ -1157.93693 &
\centering -1157.84647 &
\centering\arraybslash 14.39\\
\centering 0.01 &
\centering 181.11 &
{\centering -1157.92120\par}

~
 &
\centering -1157.83951 &
{\centering 14.46\par}

~
\\
\centering 0.015 &
\centering 182.86 &
\centering -1157.89693 &
\centering -1157.82849 &
\centering\arraybslash 14.53\\
\centering 0.02 &
\centering 184.82 &
\centering -1157.86467 &
\centering -1157.81378 &
\centering\arraybslash 14.61\\\hline
\end{supertabular}
\end{center}
\end{table}

\begin{table}[p]
\caption{Structural and electronic data for the strained/unstrained $\mathrm{GCN}-\mathrm{Fe}$ systems. The binding energies $E_b$, the averaged bond length between $\mathrm{Fe}$ atom and N atom, averaged bond length linking the heptazine, the  N-C-N angle and Fe height are refer as $d_{\mathrm{Fe-N}}$, and $d_1/d_2$, and $\theta ,h$ respectively. The charge transfer, magnetic moment per unit cell and per Fe atom, electronic character of the $\mathrm{GCN}-\mathrm{Fe}$ system and the electronic character are refer as $Q$, $M_{\mathrm{cell}}$, $M_{\mathrm{Fe}}$, $\mathrm{EC}$ respectively. M stands for metal while SC represents semiconducting.}
\label{table2}
\begin{center}
\begin{supertabular}{m{0.4in}m{0.41in}m{0.48in}m{0.6in}m{0.53375983in}m{0.3in}m{0.5in}m{0.37615985in}m{0.35in}m{0.3in}}
\hline
\\
\centering Strain &
{\centering $E_b$\par}
\centering (eV) &
{\centering $d_{\mathrm{Fe}-N}$\par}
\centering ({\AA}) &
{\centering $d_1/d_2$\par}
\centering ({\AA}) &
{\centering $\theta $\par}
\centering ($^{\circ}$) &
{\centering $h$\par}
\centering ({\AA}) &
{\centering $Q$\par}
\centering (electrons) &
{\centering $M_{\mathrm{Fe}}$\par}
\centering (\textit{{\textmu}}\textsubscript{B}) &
{\centering $M_{\mathrm{cell}}$\par}
\centering (\textit{{\textmu}}\textsubscript{B}) &
\centering\arraybslash $\mathrm{EC}$\\\hline
\centering 0\% &
\centering 2.80 &
\centering 2.37 &
\centering 1.48/1.51 &
\centering 116.46 &
\centering 0.01 &
\centering 0.73 &
\centering 3.80 &
\centering 3.96 &
\centering\arraybslash M\\
\centering 1\% &
\centering 2.49 &
\centering 2.40 &
\centering 1.51/1.53 &
\centering 116.72 &
\centering 0.00 &
\centering 0.73 &
\centering 3.82 &
\centering 3.95 &
\centering\arraybslash -\\
\centering 2\% &
\centering 1.72 &
\centering 2.43 &
\centering 1.53/1.56 &
\centering 116.99 &
\centering 0.01 &
\centering 0.74 &
\centering 3.83 &
\centering 3.95 &
\centering\arraybslash -\\
\centering 3\% &
\centering 0.54 &
\centering 2.47 &
\centering 1.57/1.60 &
\centering 117.22 &
\centering 0.05 &
\centering 0.74 &
\centering 3.84 &
\centering 3.94 &
\centering\arraybslash -\\
\centering 4\% &
\centering -0.98 &
\centering 2.48 &
\centering 1.59/1.65 &
\centering 117.67 &
\centering 0.00 &
\centering 0.74 &
\centering 3.85 &
\centering 3.93 &
\centering\arraybslash -\\
\centering 5\% &
\centering -2.20 &
\centering 2.45 &
\centering 2.61/1.46 &
\centering 118.47 &
\centering 0.10 &
\centering 0.75 &
\centering 3.84 &
\centering 3.95 &
\centering\arraybslash SC\\\hline
\end{supertabular}
\end{center}
\end{table}

\begin{figure*}[t]
{\centering  \includegraphics[width=6.5in,height=2.4in]{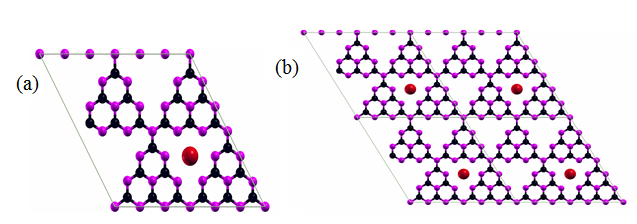} \par}
   \includegraphics[width=5.2in,height=2.8in]{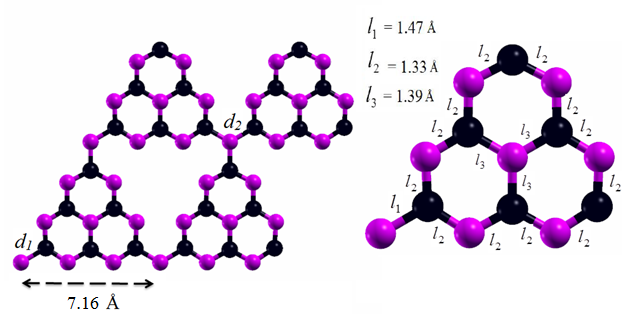} 
{\centering
\textbf{(c)}
\par}
{\centering                \includegraphics[width=4.7728in,height=3.01in]{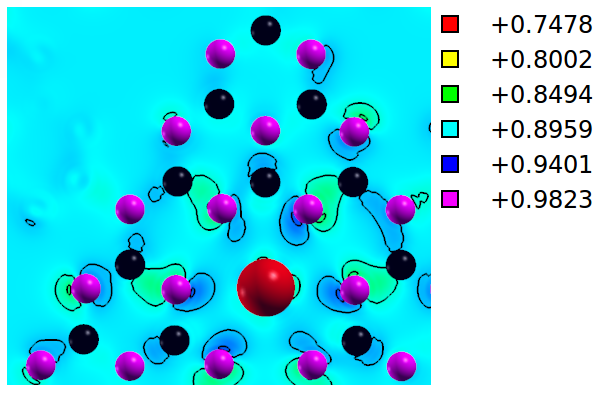} \par}
{\centering
\textbf{(d)}
\par}
\end{figure*}

\begin{figure}[t]
{\centering  \includegraphics[width=5.0in,height=0.6665in]{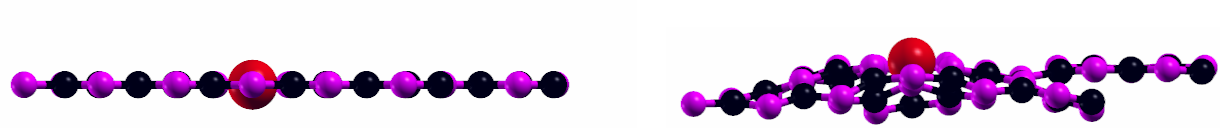} \par}
{\centering
\textbf{(e)}
\par}
 \includegraphics[width=5.0in,height=1.5in]{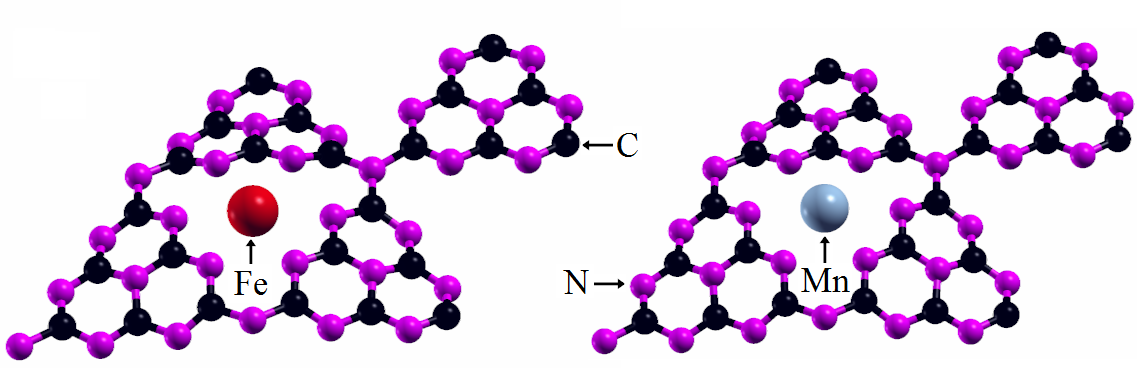} 
{\centering
\textbf{(f)}
\par}
\caption{(a) Top view of a supercell comprised of $2\times 2$ heptazine unit cell, with an embedded Fe atom. This $2\times 2$ supercell is the one that was used as the input in our calculation{. (b). }Top view of a supercell made of $4\times 4$ heptazine unit cell with embedded Fe atoms. The distance between the Fe atoms in is approximately 14.6 $\text{{\AA}}${. (c). \textbf{Left: }}Top view of a supercell comprised of $2\times 2$ heptazine unit cell. The atomic symbols and the geometric parameters are depicted in the figure. Atoms in dotted circles are the nitrogen atoms. \textbf{Right}: heptazine unit cell which is made of 3 triazine rings. {(d). }Top view plot of the charge-density difference of a supercell comprised of $2\times 2$ heptazine unit cell with embedded $\mathrm{Fe}$ atoms. Charge depletion and accumulation in a.u. are depicted by colors scale. Carbon atoms are in black. Atoms dotted with 4-point starts are the nitrogen atoms in pink color. {(e)}.\textbf{ Left}: Side view of the optimized buckled $2\times 2\times 1$ structure of $\mathrm{GCN}$ with embedded $\mathrm{Fe}$ atom without applied perpendicular electronic field. \textbf{Right}: Side view of the optimized buckled $2\times 2\times 1$ structure of $\mathrm{GCN}$ with embedded Fe atom under applied perpendicular electronic field. {(f)}.\textbf{Left}: Top view of the optimized buckled $2\times 2\times 1$ structure of $\mathrm{GCN}$ with embedded $\mathrm{Mn}$ atom under applied perpendicular electronic field. \textbf{Right}: Top view of the optimized buckled $2\times 2\times 1$ structure of $\mathrm{GCN}$ with embedded Fe atom under applied perpendicular electronic field.}
\label{fig1}
\end{figure}

\begin{figure}[!p]
{\centering
\textbf{(a)} \includegraphics[width=2.8228in,height=2.2398in]{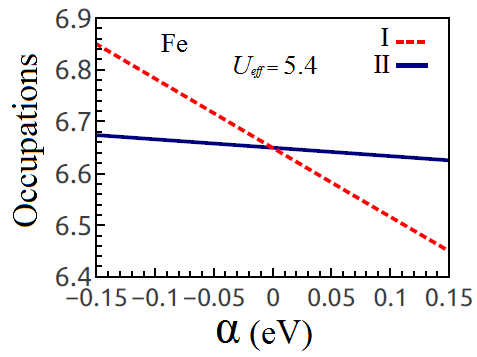} 
\par}
{\centering
 \includegraphics[width=6.5in,height=4.1in]{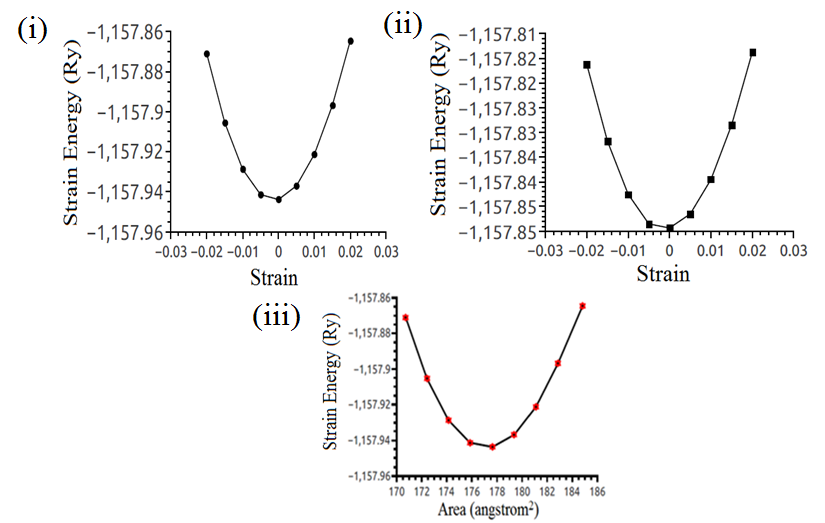} \textbf{(b)}
\par}
{\centering
\caption{(a) Linear response of d orbital occupations as a function of potential shift $\alpha$. (b) Total energy (Ry) vs. area $(\mathrm{Angstrom}^2$) of the $\mathrm{GCN}-\mathrm{Fe}$ system for bulk modulus calculations.}}
\label{fig2}
\end{figure}

\begin{figure}[!p]
\textbf{(a)} \includegraphics[width=2.4in,height=1.91in]{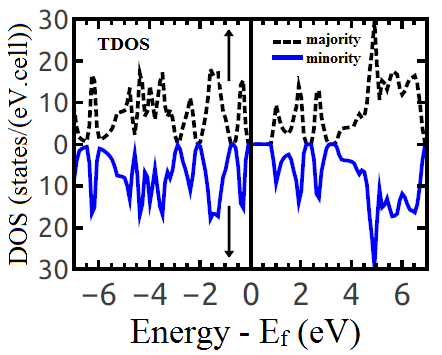} \textbf{(b)} \includegraphics[width=2.35in,height=1.9in]{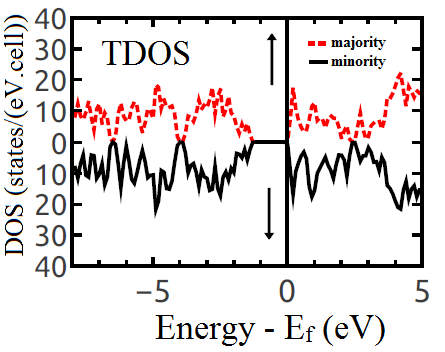} 
\textbf{(c)} \includegraphics[width=2.45in,height=1.9in]{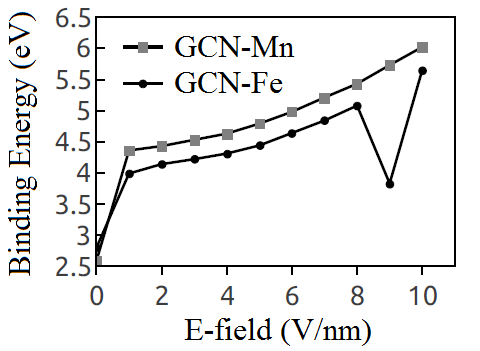} \textbf{(d)} \includegraphics[width=2.45in,height=1.9in]{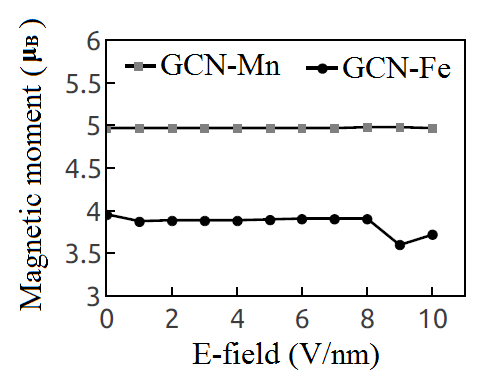} 
\caption{(a). TDOS of pristine GCN with an arrow indicating spin up and spin down directions. There is an indirect band gap in the vicinity of the Fermi level. {(b).} TDOS with an arrow indicating spin up and spin down directions for $s=5$ tensile strain $\mathrm{GCN}-\mathrm{Fe}$ system. {(c). }Dependence of binding energy on the applied electric field for $\mathrm{GCN}-\mathrm{Fe}$ and $\mathrm{GCN}-\mathrm{Mn}$ systems. {(d).} Dependence of magnetic moment per unit cell on the applied electric field for $\mathrm{GCN}-\mathrm{Fe}$ and $\mathrm{GCN}-\mathrm{Mn}$ systems.}
\label{fig3}
\end{figure}
 
\begin{figure*}[p]
{\centering
(a) \includegraphics[width=2.4in,height=1.9in]{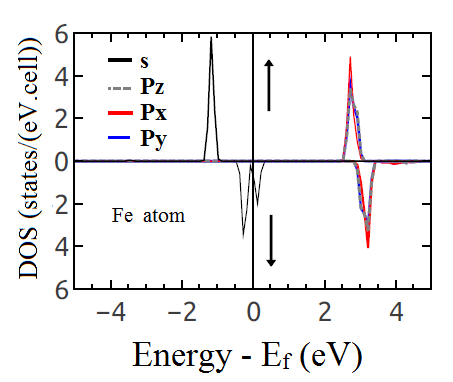} (b) \includegraphics[width=2.4in,height=1.9in]{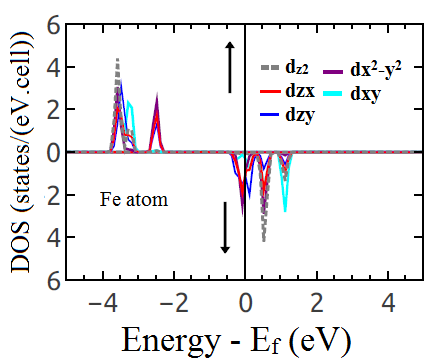} 
\par}
{(c)} \includegraphics[width=2.4in,height=1.6in]{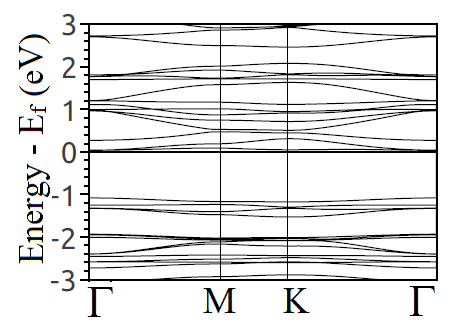}  {(d)} \includegraphics[width=2.4in,height=1.6in]{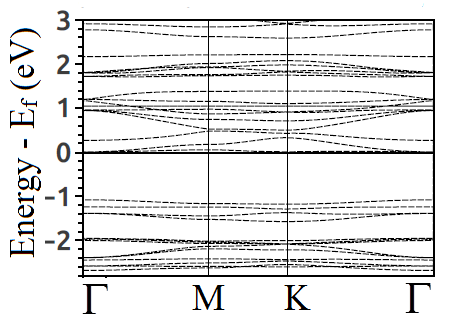}
\end{figure*}

\begin{figure}[p]
{\ \ \ \ \ \ \ \ \ \ \ \ \ \ \ \ \  Majority spin state} \ \  \ \ \ \ \ \ \ \ \ \ \ \ \ \ \ \ \ \ \ \ {Minority spin state}  \\\ {(e)} \includegraphics[width=2.4in,height=1.7in]{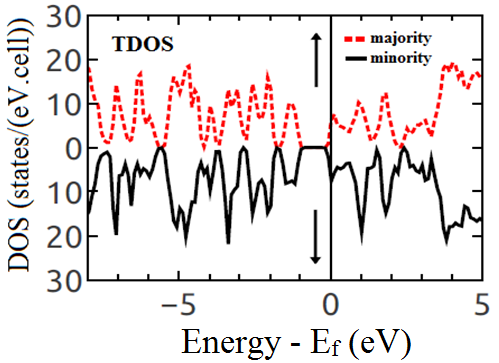} 
{f(i)} \includegraphics[width=2.4in,height=1.7in]{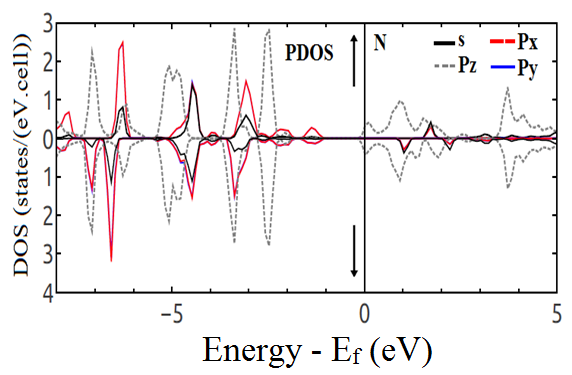}
\\
\\
{f(ii)} 
\includegraphics[width=2.3in,height=1.7in]
{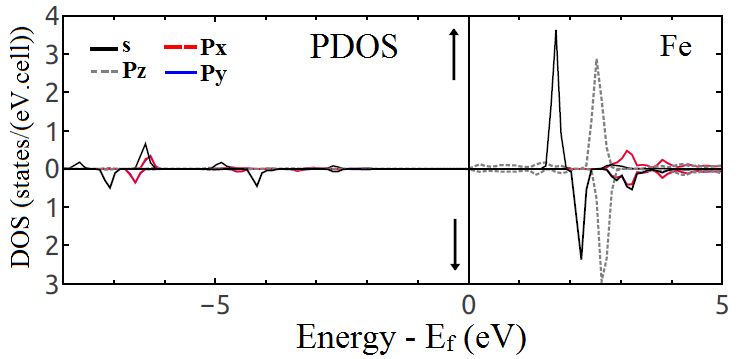} {f(iii)} \includegraphics[width=2.3in,height=1.7in]
{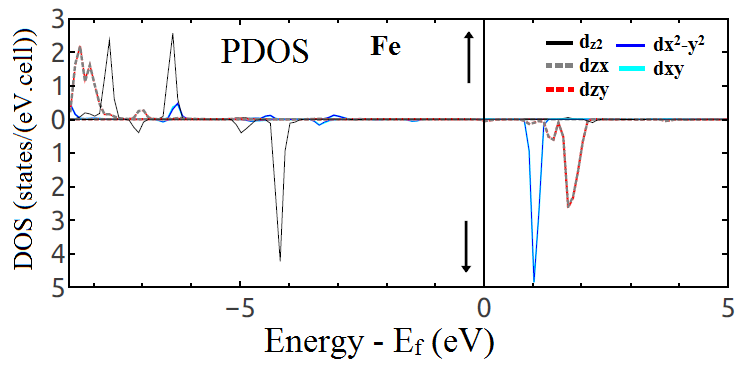}
\caption{Projected density of state ($\mathrm{PDOS}$) for free $\mathrm{Fe}$ atom; spin-polarized electronic band structure and the corresponding TDOS and PDOS for unstrained ($s=0$) $\mathrm{GCN}-\mathrm{Fe}$ system. {(a)} $\mathrm{PDOS}$ of free $\mathrm{Fe}$ atom for s, p orbitals. (b) $\mathrm{PDOS}$ of free $\mathrm{Fe}$ atom for d orbitals. {(c)} Spin up band structure of $\mathrm{GCN}-\mathrm{Fe}$ system. {(d)} Spin down band structure of $\mathrm{GCN}-\mathrm{Fe}$ system. {(e)} TDOS of the $\mathrm{GCN}-\mathrm{Fe}$ system with an arrow indicating spin up and spin down directions. {(f)} Projected density of states ($\mathrm{PDOS}$) with an arrow indicating spin up and spin down directions for (i) $\mathrm{sp}$ like-orbital of the sum of 6 edge N atoms (ii) $\mathrm{sp}$-like orbitals of the Fe atom in the $\mathrm{GCN}-\mathrm{Fe}$ system (iii) d-like orbitals of the  Fe atom in the $\mathrm{GCN}-\mathrm{Fe}$ system respectively.}
\label{fig4}
\end{figure}

\begin{figure}[!p]
{a(i)} \includegraphics[width=2.3in,height=1.7in]{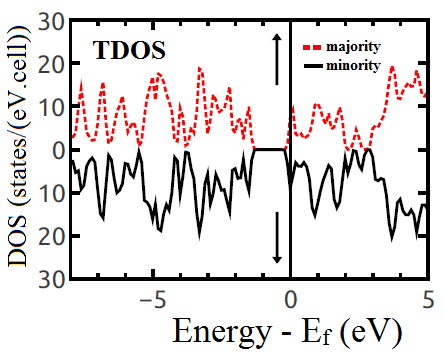} {a(ii)}  \includegraphics[width=2.3in,height=1.7in]{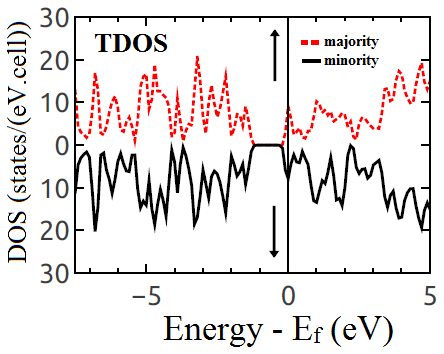} 
\ \ \ \ \ \ \ \ \ \ \ \ \ \ \ \ \ \ {GCN-Fe} 1.0 V/nm \ \ \ \ \ \ \ \ \ \ \ \ \ \ \ \ \ \ \ \ \ \ \ \ \ \ \ \ \ \ \ \ {GCN-Fe }5.0 V/nm 
\\
\\
{a(iii)} \includegraphics[width=2.3in,height=1.7in]{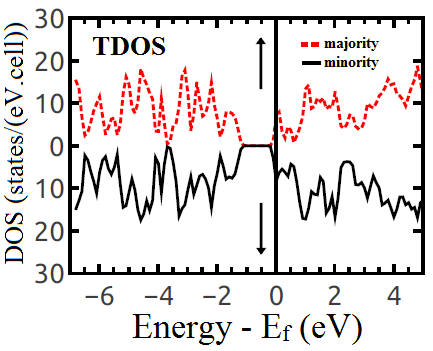} {b(i)} \includegraphics[width=2.3in,height=1.7in]{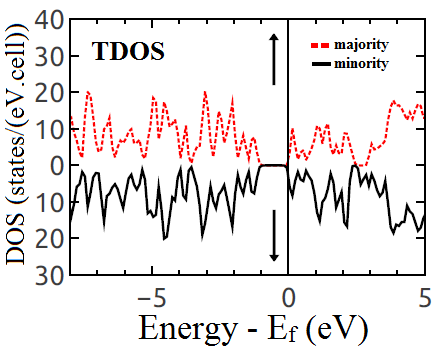} 
{{GCN-Fe} 10.0 V/nm {\ \ \ \ \ \ \ \ \ \ \ \ \ \ \ \ \ \ \ \ \ \ \ \ \ \ \ \ \ \ \ \ GCN-Mn} 1.0 V/nm}
\\
\\
{b(ii)} \includegraphics[width=2.3in,height=1.6in]{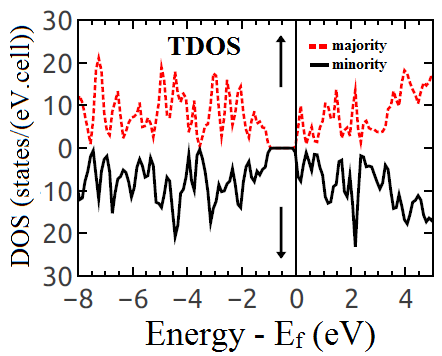} {b(iii)} \includegraphics[width=2.3in,height=1.6in]{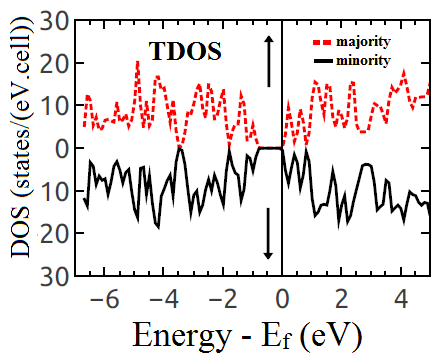} 
\ \ \ \ \ \ \ \ \ \ \ \ \ \ \ {GCN-Mn} 5.0 V/nm { \ \ \ \ \ \ \ \ \ \ \ \ \ \ \ \ \ \ \ \ \ \ GCN-Mn} 10.0 V/nm
\caption{TDOS and PDOS with an arrow indicating spin up and spin down directions of $\mathrm{GCN}-\mathrm{Fe}$ and $\mathrm{Mn}-\mathrm{FeGCN}$ systems under applied electronic field. {(a)} TDOS with an arrow indicating spin up and spin down directions of $\mathrm{GCN}-\mathrm{Fe}$ system under electronic field of magnitude (i) 1.0 $\mathrm{V/nm}$ (ii) 5.0 $\mathrm{V/nm}$ and (iii) 10.0 $\mathrm{V/nm}$ respectively. {(b).} The spin-polarized TDOS for $\mathrm{GCN}$ with embedded $\mathrm{Mn}$ atom under applied electronic field of magnitude (i) 1.0 $\mathrm{V/nm}$ (ii) 5.0 $\mathrm{V/nm}$ and (iii)  10.0 $\mathrm{V/nm}$ respectively.}
\label{fig5}
\end{figure}

\end{document}